\documentclass[12pt]{article}
\usepackage{geometry}
\usepackage{color}
\usepackage{graphics}
\usepackage{epsfig}

\makeatletter

\newcommand{\noun}[1]{\textsc{#1}}
\let\SF@@footnote\footnote
\def\footnote{\ifx\protect\@typeset@protect
    \expandafter\SF@@footnote
  \else
    \expandafter\SF@gobble@opt
  \fi
}
\expandafter\def\csname SF@gobble@opt \endcsname{\@ifnextchar[
  \SF@gobble@twobracket
  \@gobble
}
\edef\SF@gobble@opt{\noexpand\protect
  \expandafter\noexpand\csname SF@gobble@opt \endcsname}
\def\SF@gobble@twobracket[#1]#2{}

 \usepackage{verbatim}

\newcommand{\pdir}{p\kern -5.2pt\raise 0.2ex\hbox {/}}
\newcommand{\vdir}{v\kern -5.75pt\raise 0.15ex\hbox {/}}
\newcommand{\kdir}{k\kern -5.75pt\raise 0.15ex\hbox {/}}
\newcommand{\epsdir}{\epsilon\kern -5.0pt\raise 0.15ex\hbox {/}}
\newcommand{\bvdir}{\bar{v}\kern -5.75pt\raise 0.15ex\hbox {/}}
\newcommand{\Ddir}{D\kern -7.75pt\raise 0.20ex\hbox {/}}
\newcommand{\ldir}{l\kern -5.0pt\raise 0.2ex\hbox{/}}
\newcommand{\varepsdir}{\varepsilon\kern -5.5pt\raise 0.15ex\hbox{/}}

\newcommand{\etc}{{\it etc}}
\makeatother

\begin{document}
\thispagestyle{empty} 
{\par\raggedleft BUHEP-00-2\par}
{\par\raggedleft FTUV/99-69 IFIC/99-73\par}
{\par\raggedleft ORSAY/00-18\par}
{\par\raggedleft ROME1-1289/99\par}

\vskip 0.5cm

{\par\centering \textbf{\textcolor{blue}{\Huge NNLO Unquenched 
Calculation }}\\
\textbf{\textcolor{blue}{\Huge of the b Quark Mass.}}\Huge \par}

{\par\centering \vskip 0.75 cm\par}

{\par\centering \textcolor{red}{\Large V.~Gim\'enez}\\
\vskip 0.25cm\par}

{\par\centering \textit{Dep.~de F\'{\i}sica Te\`orica and IFIC, Univ.~de 
Val\`encia,}\\
\textit{Dr.~Moliner 50, E-46100, Burjassot, Val\`encia, 
Spain}\textsf{\textit{.}}\\
\vskip 0.5cm\par}

{\par\centering \textcolor{red}{\Large L.~Giusti}\\
\vskip 0.25cm\par}

{\par\centering \textit{Department of Physics, Boston University}\\
\textit{Boston, MA 02215 USA}.\\
\vskip 0.5cm\par}

{\par\centering \textcolor{red}{\Large G.~Martinelli$^{\dagger}$}\\
\vskip 0.25cm\par}

{\par\centering \textit{Universit\'e de Paris Sud, LPT B\^at.~210, }\\
\textit{91405 Orsay-Cedex, France}.\\
\vskip 0.5cm\par}

{\par\centering \textcolor{red}{\Large F.~Rapuano}\\
\vskip 0.25cm\par}

{\par\centering \textit{Dip.~di Fisica, Univ.~di Roma \char`\"{}La 
Sapienza\char`\"{}
and}\\
\textit{INFN, Sezione di Roma, P.le A.~Moro 2, I-00185 Roma, 
Italy.}\par}
\vskip 0.75cm
\hrule

\begin{abstract}
\textcolor{blue}{By combining the first unquenched lattice computation of the B-meson 
binding energy and the two-loop contribution to the lattice HQET
residual mass, we determine the \( \overline{{MS}} \) \( b \)-quark  mass,
\( \overline{m}_{b}(\overline{m}_{b}) \). The inclusion of the 
two-loop corrections is essential to extract \( \overline{m}_{b}(\overline{m}_{b}) \) with 
a precision of  \( {\cal O}(\Lambda^{2}_{QCD}/m_{b}) \), which is the uncertainty due to
the renormalon singularities in the perturbative series of the
residual mass.  Our best estimate is \( 
\overline{m}_{b}(\overline{m}_{b})\, =\, (\, 4.26\, \pm \, 0.09\, )\, 
{\rm GeV} \), where we have combined the different errors in quadrature.
A detailed discussion of the systematic errors contributing to  the final number  is presented.
Our results have been obtained on a sample of \( 60 \) lattices of 
size \( 24^{3}\times 40 \) at \( \beta =5.6 \), using the Wilson action for light 
quarks and the lattice HQET for the \( b \)  quark, at two  values of the sea quark masses. The
quark propagators have been computed using the unquenched links 
generated by the T\( \chi  \)L Collaboration.}
\end{abstract}
\hrule

\vskip 0.5cm

{\par\centering PACS: 11.15.Ha; 12.38.Gc; 12.39.Hg; 12.15.Ff and 14.65.Fy\par}
\vskip 0.5 cm 
{\par\centering $^{\dagger}$ On leave of absence from Dip. di Fisica, Univ. di Roma \char`\"{}La 
Sapienza\char`\"{} }
\newpage
\clearpage
\setcounter{page}{1}

\section{Introduction}

Quark masses are fundamental parameters of QCD that  cannot be  determined
by theoretical considerations only and, due to the confinement of  quarks inside
hadrons,  cannot be  directly measured.  Quark masses can, however, be 
introduced as short-distance effective couplings. As such,  they are scale
and scheme dependent quantities, the  values
of which  depend on the adopted definition. Nonetheless, quark masses 
are very important for phenomenology since they enter many theoretical predictions of 
physical quantities such as   CKM matrix elements,  $b$-hadron  inclusive semileptonic 
decays, total widths,  \etc. This is the reason why, in the last 
years, much effort has been devoted to  accurately determine their values. 

It is useful to classify  quarks into two classes: light quarks, 
the masses of which are lower or of the order of \( \Lambda _{QCD} \) (the \( u \), \( d 
\) and \( s \) quarks are light), and heavy quarks, with masses larger than
\( \Lambda _{QCD} \) ( \( b \) and \( t \) quarks are heavy and, to  some extent,
the charm quark \( c \)  too). Light-quark masses are 
extracted from hadron spectroscopy using lattice QCD simulations~\cite{latms} and also  QCD sum
 rules and \( \tau  \) decay data~\cite{taums}.  Heavy-quark masses can be
extracted from the properties of hadrons containing heavy quarks: the
\( B \)-meson spectrum from the lattice HQET~\cite{oursmb,NNLO}, the \( 
\Upsilon  \) (or
$J/\psi$) spectrum with lattice NRQCD~\cite{nrqcd,nrqcd2} or QCD Sum Rules 
~\cite{narison}--\cite{penin}
and mass effects in 3--jets \( b\bar{b}g \) events 
~\cite{arcadi,delphi}. 

In this paper, we present the first unquenched HQET lattice 
calculation of the \( b \) quark mass. The  idea~\cite{oursmb} (see 
also~\cite{methodtheo}) is to combine the  HQET unquenched lattice computation of the \( B 
\)-meson binding energy~\cite{pisa} with the recent next-to-next-to-leading 
( NNLO ) perturbative  calculation of the matching of the  \( \overline{MS} \) 
quark mass to its lattice HQET counterpart~\cite{NNLO}. We  stress that
both  unquenched  simulations and NNLO matching are necessary
ingredients to improve the accuracy of previous  results~\cite{oursmb}. The former
is necessary to control potentially large vacuum-polarization 
contributions to the \( B \)-meson propagator. The latter is crucial to reduce  
renormalon ambiguities in the continuum-lattice matching~\cite{NNLO}. 
After a careful analisys of the systematics errors, our best result
is
\textcolor{red}{
\[
\overline{m}_{b}(\overline{m}_{b})\, =\, (\, 4.26\, \pm \, 0.03\, \pm 
0.05\, \pm \, 0.07\, )\, {\rm GeV}\, , \] }
where the first error is statistical; the second is the systematic 
error  from the spread of values due to the use of different time intervals, 
fitting methods, smearing types and cube sizes for the interpolating operators, the 
dependence of the results on the mass of the sea quarks, 
the calibration of the lattice spacing and an evaluation  of the \( 
1/m_{b} \) corrections; 
the third is an estimate of the error due to the uncertainties in the values 
of \( \alpha _{s} \)  and to the effects of higher-order terms  
in eq.~(\ref{eq:masterformulaallorders}). A detailed discussion of 
the different errors can be found below.

The paper is organized as follows: in Sect.~2, we  briefly describe 
our method and give the main formulae we used; in Sect.~3, we discuss 
the  lattice computation of the binding energy;  details of the simulation
and  numerical results   are presented in 
Sect.~4, where we  also discuss
the procedure used for analyzing the unquenched lattice data; in 
Sect.~5, we carefully study the different sources of systematic errors in our results. 
Finally, in  Sect.~6,  we present our final numbers and compare them with other recent 
determinations.

\section{The method.}

The key idea to determine the \( b \)-quark mass  consists in 
matching  the  propagator in QCD to its lattice
HQET counterpart~\cite{oursmb}. As shown below, this matching allows
us to relate the pole  mass to the binding energy and to the physical 
mass of
the \( B \) meson. The renormalized \( \overline{MS} \) \( b \)-quark 
mass at a given
scale  \( \mu  \)  can then be obtained from the pole mass  by using 
perturbation
theory. In this section we briefly recall   the  formulae  relevant 
to our study.

Lattice HQET is an effective theory of QCD. The relation between
the inverse \( b \)-quark propagator in QCD, \( S^{-1} \), and its 
lattice HQET counterpart, \( {S}_{L}^{-1} \), can be written, to 
lowest  order in \( 1/m_{b} \), as
\begin{equation}
\label{eq:matching}
\left( \frac{1+\vdir }{2}\right)\, S_{P}^{-1}(p,m_{b};\mu )\, \equiv \, \left( \frac{1+\vdir }{2}\right) 
\, S^{-1}\, \left( \frac{1+\vdir }{2}\right) =\, C(\mu\, a,\alpha 
_{s})\, \left( \frac{1+\vdir }{2}\right)\, {S}^{-1}_{L}((v\cdot k)a)
\, , \end{equation}
 where \( S_{P}^{-1} \) is the projected \( b \)-quark propagator, \( 
\mu  \)
 the renormalization point, \( a \)  the lattice spacing, \( 
p=m_{b}\, v\, +\, k \)
 the momentum of the \( b \)-quark, \( v \)  its velocity and \( k \)
 the residual momentum. \( C(\mu\, a, \alpha_{s}) \) is the 
relevant 
Wilson coefficient. It contains all the mass dependence of the right 
hand side of eq.~(\ref{eq:matching}) since, by construction, the HQET 
propagator is independent of the \( b \) quark mass. It should be noticed that in 
order for the HQET to be applicable, \( k \) must satisfy  the condition \( 
|k|\ll m_{b} \). In
writing eq.~(\ref{eq:matching}) we have chosen as expansion 
parameter the quark mass  appearing in  the original propagator, namely  \( m_{b} \). 

The procedure to find \( C \) is well known: calculate the \( b 
\)-quark propagator in QCD and in the lattice HQET to a given order in 
\( \alpha _{s} \),  expand the former in inverse powers of \( m_{b} \) to a given order (lowest 
order in our case), and finally compare both expressions at a fixed scale 
$\mu$ (with $\mu \gg  \Lambda _{QCD}$) to  extract \( C(\mu\, a, \alpha _{s}) \). 
Renormalization group can then be used to evolve this function to any 
scale. 

To illustrate and clarify the key points of this strategy, we briefly 
sketch
the derivation of our master formula for the \( b \)-quark mass to \( 
O(\alpha _{s}) \)
and then we extend our equation to include higher orders in 
perturbation theory.
The inverse quark propagator in QCD can be written in the form
\begin{equation}
i\, S^{-1}(p,m_{b};\mu )\, =\, \pdir \, -\, m_{b}\, +\, \Sigma 
_{1}(p^{2},m_{b})\, +\, (\pdir \, -\, m_{b})\, \Sigma 
_{2}(p^{2},m_{b}) \, .
\end{equation}
It is very easy to calculate the self-energy form factors \( \Sigma 
_{1} \)
and \( \Sigma _{2} \) to one loop in some renormalization scheme and 
for a
fixed gauge. By writing the \( b \)-quark momentum in the \( B \) 
meson, \( p \),
as \( p=m_{b}\, v\, +\, k \) and expanding in powers of \( 1/m_{b} 
\), one finds 
\begin{eqnarray}
S_{P}^{-1}(p,m_{b};\mu ) & = & m_{b}\, -\, m_{b}^{pole}  
\label{eq:propSp}\nonumber \\
 &+& (v\cdot k)\, \left[ 1+\alpha _{s}(\mu )\, \left( \gamma 
_{D\cdot k}\, \ln \left( \frac{\mu}{-2(v\cdot k)}\right) \, +\, 
c_{2}\right) \, +\, \cdots \right]\, , 
\end{eqnarray}
 where  \( c_{2} \) is a  scheme dependent constant (the 
expression of which 
is irrelevant for our discussion)  and \( \gamma _{m} \) and \( 
\gamma _{D\cdot k} \) are
the scheme-independent one-loop anomalous dimensions of the mass and  
operator
\( D\cdot k \), respectively. 

The pole mass, \( m_{b}^{pole} \), is defined as the position of the 
pole of
the propagator \( S^{-1} \), at a given order in perturbation theory,
\begin{eqnarray}
S_{P}^{-1}(\pdir \, =\, m_{b}^{pole},m_{b};\mu ) \, =\, 0  \, .
\end{eqnarray}
To one loop, the explicit calculation of  eq.~(\ref{eq:propSp}) gives
\begin{equation}
\label{polemass}
m_{b}^{pole}\, =\, m_{b}\, \left[ 1+\alpha _{s}(\mu )\, \left( \gamma 
_{m}\, \ln \left( \frac{m_{b}}{\mu }\right) \, +\, c_{1}\right) 
\right] \, .
\end{equation}

In order to implement the matching \( 
{S}_{L}^{-1} \),  the propagator  of
 the lattice HQET, must be evaluated at the same order in perturbation 
theory
\begin{equation}
\label{eq:propSl}
{S}_{L}^{-1}((v\cdot k)a)\, =\, (v\cdot k)\, \left[ 1+\alpha 
_{s}(a)\, \left( \gamma _{D\cdot k}\, \ln \left( \frac{1}{-2(v\cdot 
k)\, a}\right) \, +\, d_{2}\right) \right] \, -\, \alpha _{s}(a)\, 
\frac{X_{0}}{a}\, +\, \cdots 
\end{equation}
where  \( d_{2} \) is a scheme dependent constant. From 
eq.~(\ref{eq:propSl}) we learn that an additive, linearly divergent mass 
term is generated on the lattice:
the so-called residual mass, \( \delta m \). 
 Inserting  eq.~(\ref{eq:propSl}) into eq.~(\ref{eq:propSp}) and 
taking into account the
expression of the pole mass in eq.~(\ref{polemass}), we obtain
\begin{equation}
\label{eq:matching2}
S_{P}^{-1}(p,m_{b};\mu )\, =\, m_{b}\, -\, m^{pole}_{b}\, +\, \alpha 
_{s}(\mu )\, \frac{X_{0}}{a}\, +\, C(\mu\, a,\alpha _{s})\, 
{S}^{-1}_{L}((v\cdot k)a) \, ,
\end{equation}
where the Wilson coefficient has the form
\begin{equation}
\label{eq:wilsoncoeff}
C(\mu\, a,\alpha _{s})\, =\, 1\, +\, \alpha _{s}(\mu )\, \left( 
\gamma _{D\cdot k}\, \ln \left(\mu\, a\right) \, +\, c_{2}\, -\, 
d_{2}\right) 
\end{equation}
and we have used the fact that the difference between \( \alpha 
_{s}(\mu ) \) and \( \alpha _{s}(a) \)
is \( {\cal O}(\alpha _{s}^{2}) \). Comparing
eq.~(\ref{eq:matching}) with eq.~(\ref{eq:matching2}),  the important
relation between the HQET expansion mass parameter \( m_{b} \) and 
the pole \( b \)-quark mass can be derived 
\begin{equation}
\label{eq:poletomb}
m_{b}^{pole}\, =\, m_{b}\, +\, \alpha _{s}(\mu )\, \frac{X_{0}}{a}\, 
\equiv \,
m_{b}\, +\, \delta m \, .
\end{equation}
To lowest order in \( 1/m_{b} \), the HQET mass formula can now be 
used to eliminate the unknown expansion parameter, \( m_{b} \), by expressing 
it in terms of the physical mass of a $b$-hadron, specifically the 
\( B \)-meson,  and the non-perturbative  binding energy, \( {\cal E} \), which is 
independent of \( m_{b} \),
\begin{equation}
\label{eq:hqetmassformula}
M_{B}\, =\, m_{b}\, +\, {\cal E}\, +\, {\cal O}(1/m_{b})
\end{equation}
Using the equation above, we get  
\begin{equation}
\label{eq:polemass}
m_{b}^{pole}\, =\, M_{B}\, -\, {\cal E}\, +\, \delta m\, +\, {\cal 
O}(1/m_{b}) \, .
\end{equation}
 Finally, the pole mass is converted into the \( \overline{MS} \) 
mass through
the well-known one-loop perturbative relation 
\begin{eqnarray}
\overline{m}_{b}(\overline{m}_{b})\,  & = & m_{b}^{pole}\, \left[ 1\, 
-\, \frac{4}{3}\, \left( \frac{\alpha _{s}(\overline{m}_{b})}{\pi 
}\right) \right] \label{eq:masterformula} \nonumber \\
 & = & \left[ M_{B}\, -\, {\cal E}\, +\, \alpha 
_{s}(\overline{m}_{b})\, \frac{X_{0}}{a}\right] \, \left[ 1\, -\, 
\frac{4}{3}\, \left( \frac{\alpha _{s}(\overline{m}_{b})}{\pi 
}\right) \right] \, +\, {\cal O}(1/m_{b}) \, .
\end{eqnarray}
We stress that $\overline{m}_{b}(\overline{m}_{b})$ is obtained from 
the non-perturbative quantity ${\cal E}$ combined with the 
perturbative calculation of lattice ($\alpha_{s} X_{0}$) and continuum
($4/3\,\alpha _{s}(\overline{m}_{b})/\pi$) coefficients.

The generalization of eq.~(\ref{eq:masterformula}) to higher orders 
is straightforward. One gets
\begin{eqnarray}
\overline{m}_{b}(\overline{m}_{b})\,  & = & m_{b}^{pole}\, \left[ 1\, 
+\, \sum ^{\infty }_{n=0}\, \left( \frac{\alpha 
_{s}(\overline{m}_{b})}{\pi }\right) ^{n+1}\, D_{n}\right] \, +\, 
{\cal O}(1/m_{b}) \label{eq:masterformulaallorders} \nonumber \\
 & = & \left[ M_{B}\, -\, {\cal E}\, +\, \sum ^{\infty }_{n=0}(\alpha 
_{s}(\overline{m}_{b}))^{n+1}\, \frac{X_{n}}{a}\right] \, \left[ 1\, 
+\, \sum ^{\infty }_{n=0}\, \left( \frac{\alpha 
_{s}(\overline{m}_{b})}{\pi }\right) ^{n+1}\, D_{n}\right] \, ,
\end{eqnarray}
which is the master equation of our analysis. \( D_{n} \) and \( 
X_{n} \)
are constants which depend on the number of flavours, \( n_{f} \), 
the masses
of the active quarks and the lattice action used for the light quarks 
(see below). 
\vskip 0.2 cm
The procedure used to calculate \( \overline{m}_{b}(\overline{m}_{b}) \)
is the following~\cite{oursmb}:

\begin{itemize}
\item compute the binding energy of the \( B \) meson in 
lattice units, \( a {\cal E} \), using the HQET for the  heavy quark and a given discretized action 
(Wilson, Alpha, Staggered) to describe the light-quark dynamics (we discuss in detail 
this computation in Sects. 3 and 4).
\item evaluate  the value of the lattice spacing, \( a \), from 
the light-hadron spectroscopy.
\item take the experimental value of the \( B \)-meson mass, \( M_{B} 
\), as input.
\item insert the values of these quantities in 
eq.~(\ref{eq:masterformulaallorders})
and obtain \( \overline{m}_{b}(\overline{m}_{b}) \) to a given order 
in perturbation theory and up to \( {\cal O}(1/m_{b}) \) corrections. 
\end{itemize}
A few  important remarks are in order at this point:

\begin{enumerate}
\item The bare binding energy \( {\cal E} \) is not a physical quantity 
since it diverges linearly as \( a\rightarrow 0 \) and needs to be subtracted~\cite{maiani,beneke}.
Since both the pole and the \( \overline{MS} \) mass are finite 
quantities, instead, the divergence of \( {\cal E} \)  is cancelled by the corresponding divergence 
of the residual mass \( \delta m \),  expressed in terms of the constants \( X_{n} \). 
\item In practice the cancellation is incomplete because 
we only know the va\-lues of few cons\-tants \( X_{n} \). Therefore, at a 
given order of the perturbative expansion, we cannot take 
the lattice spacing  too small.
\item The large-\( n_{f} \) approximation  shows that the perturbative 
series for \( \delta m \), and hence for the pole mass, suffers from
 renormalon singularities~\cite{methodtheo,beneke}.
In other words, the coefficients \( X_{n} \) are expected to grow as  \( const. \times n! \)
as \( n\, \rightarrow \, \infty  \). These singularities give rise to  ambiguities
of \( {\cal O}(\Lambda _{QCD}) \) in the sum of the perturbative 
series. A solution to this problem, which is the one adopted here, is to 
consider a short distance definition of the \(b\)-quark mass, such as the \( \overline{MS} \), \( 
\overline{m}_{b} \),  because it is free of  renormalon ambiguities 
(up to \( {\cal O}(\Lambda^{2}_{QCD}/m_{b}) \) at the order at which 
we are working).
\item In the expression of the \( \overline{MS} \) mass, a delicate 
cancellation of renormalon singularities occurs: the  renormalon of the series 
for \( \delta m \) (with coeffcients \( X_{n} \)) is cancelled by  the 
perturbative expansion of the coefficient relating  the pole and the \( \overline{MS} \) mass (with 
coeffcients \( D_{n} \)). 
\item In order to achieve the cancellation of  renormalon 
singularities in eq.~(\ref{eq:masterformulaallorders}),
the same coupling constant has to be used in the expansion of \( 
\delta m \) and in the relation between the pole mass and the \( \overline{MS} \) 
mass~\cite{methodtheo,NNLO}. For this reason, although we work at a 
fixed order of perturbation theory, we believe that the most 
reasonable choice is to expand  both  the continuum and lattice series using the same coupling constant.
\end{enumerate}
From the discussion above,  it is clear that  the precision of our 
results for the \( b \)-quark mass at  given \( ({\cal E},\, a) \)  is limited by the number of terms calculated 
in  lattice (\( X_{n} \)) and in  continuum (\( D_{n} \)) perturbation 
theory.
The relation between the pole and the \( \overline{MS} \) mass has 
been obtained to \( {\cal O}(\alpha _{s}^{2}) \) by Gray \emph{et al.}~\cite{gray90} and,  recently,
to \( {\cal O}(\alpha _{s}^{3}) \) by refs.~\cite{chetyrkin,melni}
\begin{eqnarray}
D_{0} & = & -\frac{4}{3}\, , \nonumber \\
D_{1} & = & -11.6656\, +\, 1.0414\, \sum ^{n_{f}}_{i=1}\, \left[ 
1-\frac{\overline{m}_{i}}{\overline{m}_{b}}\right] \, , \nonumber \\
D_{2} & = & -157.116 \, +\, 23.8779 \, n_{f}\, -0.6527\, 
n_{f}^{2} \, . \label{eq:valorsdeD} 
\end{eqnarray}
Note that the three-loop correction has been evaluated 
with massless quarks. 

As for the residual mass, \( \delta m \), it can be expressed in 
terms of the
bare lattice coupling, \( \alpha _{0} \), as
\begin{equation}
\label{eq:deltamalpha0}
\delta m\, =\, \sum ^{\infty }_{n=0}(\alpha _{0})^{n+1}\, 
\frac{\overline{X}_{n}}{a}
\end{equation}
 The constant \( \overline{X}_{0} \) is simply \( 
\overline{X}_{0}=X_{0} \)
given by the  three-dimensional integral
\begin{equation}
\label{eq:X0}
X_{0}\, =\, C_{F}\, \frac{1}{8\pi ^{2}}\, \int ^{\pi }_{-\pi }\, 
d^{3}k\, \frac{1}{2\, \sum _{i=1}^{3}\, \sin ^{2}(k_{i}/2)}\, =\, 
2.1173
\end{equation}
 where \( C_{F}=(N^{2}-1)/2N \) and \( N \) is the number of colors. 
Martinelli and Sachrajda have performed  the calculation of \( \overline{X}_{1} \) by extracting \( 
\delta m \) from the exponential decrease of the expectation  value of
large Wilson loops with the perimeter~\cite{NNLO}. More recently,
 Burgio \emph{et al.}  have 
obtained a preliminary estimate of \( \overline{X}_{2} \) from large Wilson 
loops computed with the Numerical Stochastic Perturbation Theory (NSPT) on a \( 
24^{4} \) lattice in the quenched approximation~\cite{direnzo99}.
In summary, the results are
\begin{eqnarray}
\overline{X}_{0} & = & 2.1173\nonumber \\
\overline{X}_{1} & = & 11.152\, +\, n_{f}\, (-\, 0.282\, +\, 0.035\, 
c_{SW}-0.391\, c_{SW}^{2})\nonumber \\
\overline{X}_{2} & = & 73.5(9.2)\label{eq:resultsforbarX} 
\end{eqnarray}
where the value of the coefficient \( c_{SW} \) depends on the lattice
fermions used in the simulation: for Wilson fermions  \( c_{SW}=0 
\), for Clover-SW tree-level improved fermions \( c_{SW}=1 \) and for 
the non-perturbatively improved ones, \( c_{SW} \) depens on \( \beta  \) (see~\cite{NNLO} 
for details).
The numerical value of \( \overline{X}_{2} \) has been obtained in 
the quenched approximation (\( n_{f}=0 \)) and thus it does not include fermion-loop  effects. 

The next step is to express \( \delta m \) in terms of the \( 
\overline{MS} \) coupling \( \alpha _{s} \), i.e. to calculate the coefficients \( 
X_{n} \) from the \( \overline{X}_{n} \) of eq.~(\ref{eq:resultsforbarX}). 
The relation between \( \alpha _{s} \) and \( \alpha _{0} \) can be written, to \( 
{\cal O}(\alpha _{0}^{3}) \), as: 
\begin{equation}
\label{eq:alphaMSalpha0}
\alpha _{s}\left( \mu \right) \, =\, \alpha _{0}\, +\, 
d_{1}(\mu a)\, \alpha _{0}^{2}\, +\, d_{2}(\mu a)\, \alpha _{0}^{3}\, 
+\, \cdots 
\end{equation}
The pure gauge contributions to \( d_{1} \) and \( d_{2} \)  have been calculated in ref.~\cite{hasen2}
and~\cite{luscher95} respectively. The quark  contribution
to \( d_{1} \) for Wilson fermions can be found in~\cite{kawai81} and for
improved fermions with generic \( c_{SW} \)  in~\cite{NNLO} (see 
also~\cite{sint}). Unfortunately, the quark contribution to
the two-loop coefficient \( d_{2} \) is still unknown.  This 
calculation is necessary if $X_{2}$ is used to obtain the N$^{3}$LO  
mass in the unquenched case. So far we have
\begin{eqnarray}
d_{1}(x ) & = & -\frac{\beta _{0}}{2\pi }\, \ln (x)\, -\, 
\frac{\pi }{2\, N}\, +\, 2.13573\, N\, 
\nonumber \\ &+ &\, n_{f}\, (-0.08413\, +\, 
0.0634\, c_{SW}\, -\, 0.3750\, c_{SW}^{2})\\
d_{2}(x ) & = & d_{1}(x )^{2}-\frac{17N^{2}}{12\pi ^{2}}\, \ln 
(x )\, +\, \frac{3\pi ^{2}}{8\, N^{2}}\, -\, 2.8626216\, +\, 
1.249116\, N^{2} \nonumber 
\end{eqnarray}
where \( \beta _{0}=\frac{1}{3}\, (11\, N\, -\, 2\, n_{f}) \) and 
 $d_{2}$ is given for  $n_{f}=0$. By 
inverting eq.~(\ref{eq:alphaMSalpha0}) for \( \alpha _{0} \) and inserting it 
in eq.~(\ref{eq:deltamalpha0}), for \( N=3 \) we get the values of the constants \( 
X_{0,1,2} \):
\begin{eqnarray}
X_{0} & = & 2.1173\nonumber \\
X_{1} & = & -1.30533\, +\, 3.70677\, \ln (\overline{m}_{b}a)\nonumber 
\\
 & + & n_{f}\, (-0.103872-0.224653\, \ln 
(\overline{m}_{b}a)-0.0992368\, c_{SW}+0.402988\, 
c_{SW}^{2})\nonumber \\
X_{2} & = & \overline{X}_{2}\, +\, 6.48945\, (-3.57877\, +\, \ln 
(\overline{m}_{b}a))\, (3.29596\, +\, \ln 
(\overline{m}_{b}a)) \label{eq:resultsforX} 
\end{eqnarray}
Since the value of \( {X}_{2} \) is 
preliminary,   we cannot really  use it to obtain our final result.
We will only show that, even in the quenched case,
the result of ref.~\cite{direnzo99} is not precise enough to obtain a useful
information.

\section{Lattice computation of \protect\( {\cal E}\protect \).}

The bare binding energy of the \( B \) meson, \( {\cal E} \), is 
measured
on the lattice by studying the large-time behaviour of the \( B \) 
meson propagator~\cite{ape}. The \( b \)-quark is described by the discretized 
version of the
HQET,
\begin{equation}
{\cal L}_{HQET}\, =\, \bar{b}(x)\, D_{4}\, b(x)\, , \end{equation}
with the covariant derivative defined as
\begin{equation}
D_{4}\, b(x)\, =\, U_{\mu }(x)\, b(x+\hat{\mu })\, -\, b(x) \, , 
\end{equation}
 where \( \hat{\mu } \) indicates the \( \mu  \)-direction and \( 
U_{\mu }(x) \)
is the link variable between the lattice sites \( x \) and \( 
x+\hat{\mu } \).
Light quarks, \( q \), are simulated with some fermion action, in our  case
the Wilson action.

It is well known that correlation functions involving   heavy quarks 
suffer from a large contamination from higher-mass excitations  to the  lightest-state contribution,
to which we are  interested in. In order to improve the isolation 
of the lightest  state, we use cube and double-cube smeared axial-current
operators of size \( L_{s}  \), as interpolating operators of the B 
meson~\cite{ape}:
\begin{eqnarray}
A^{L}_{\mu }(x) & = & \bar{b}(x)\, \gamma _{\mu }\, \gamma _{5}\, 
q(x)\, , \nonumber \\
A^{S}_{\mu }(x) & = & \sum ^{L_{s}}_{i}\, \bar{b}(x_{i})\, \gamma 
_{\mu }\, \gamma _{5}\, q(x)\, , \\
A^{D}_{\mu }(x) & = & \sum ^{L_{s}}_{i,j}\, \bar{b}(x_{i})\, \gamma 
_{\mu }\, \gamma _{5}\, q(x_{j}) \, .\nonumber
\end{eqnarray}
From the above operators we construct the two-point correlation functions
\begin{equation}
C_{L_{s}}^{RR'}(t)\, =\, \sum _{\vec{x}}\, \langle 0|\, 
A^{R}_{4}(\vec{x},t)\, A^{R'\, \dag }_{4}(\vec{0},0)\, |0\rangle  \, ,\end{equation}
where \( R,R' \) stands for \( L \) (Local), \( S \) (Single smeared) 
and \( D \) (Double smeared) interpolating operators. The correlation 
functions are computed after rotating the links in the Coulomb gauge.

At large time distances, \( C^{RR'}_{L_{s}}(t) \) behaves as:
\begin{equation}
\label{eq:largetimeofC}
C_{L_{s}}^{RR'}(t)\, \longrightarrow \, Z_{L_{s}}^{R}\, 
Z_{L_{s}}^{R'}\, e^{-{\cal E}\, t}\, , 
\end{equation}
where
\begin{equation}
Z_{L_{s}}^{R}\, =\, \frac{1}{\sqrt{2\, M_{B}}}\, \langle 0|\, 
A^{R}_{4}(\vec{0},0)\, |B\rangle \, .\end{equation}
 By fitting the large time behaviour of \( C^{RR'}(t) \) to  eq.~(\ref{eq:largetimeofC}),
the bare binding energy can be extracted. Due to 
contamination from excited states, the interpolating operators couple to the ground 
state in different ways so that the actual value of \( {\cal E} \) has some dependence 
on the cube size and the smearing type used in the analysis. To obtain our best estimate, we 
compare different methods and account the mixing as a systematic  
effect contributing to the final error.

We first use the \textcolor{red}{Best Cube Method (BCM)}:  we base our results 
on the best cube, defined as the one which yields the largest and flattest 
effective-mass plateau~\cite{ape}. In practice, we proceed as the  following example illustrates. 
Consider the double smeared interpolating operators. For \( t\geq 
t_{min} \) and for each \( L_{s} \), we fit \( C_{L_{s}}^{DD}(t) \) and \( 
C_{L_{S}}^{LD}(t) \) to a single-state propagator (\ref{eq:largetimeofC}). \( t_{min} 
\)  is taken as the time at which we start observing a  plateau for
both the effective mass \( \Delta E_{L_{s}}^{DD} \) and the ratio \( 
{\rm R}_{L_{s}}^{DL/DD} \) (we call  BC DL/DD method the case where the smearing is in the sink) 
or, alternatively,
for the effective mass \( \Delta E_{L_{s}}^{LD} \) and the ratio \( 
{\rm R}^{LD/DD}_{L_{s}} \) (we call   BC LD/DD method the case where the smearing is in the 
origin). Effective masses and ratios are defined by
\begin{eqnarray}
\Delta E^{RR'}_{L_{s}}(t) & = & \ln \left( 
\frac{C_{L_{s}}^{RR'}(t)}{C_{L_{s}}^{RR'}(t+1)}\right) \, \rightarrow 
\, {\cal E} \, , \nonumber\\
{\rm R}_{L_{s}}^{RR'/NN'}(t) & = & 
\frac{C^{RR'}_{L_{s}}(t)}{C_{L_{s}}^{NN'}(t)}\, \rightarrow \, 
\frac{Z_{L_{s}}^{R}\, Z_{L_{s}}^{R'}}{Z^{N}_{L_{s}}\, Z_{Ls}^{N'}}\, .
\end{eqnarray}
For \( t\geq t_{min} \), the ground state is assumed to have been 
isolated.
The next step is to combine the exponential fit for \( 
C_{L_{s}}^{DD}(t) \)
(\( C_{L_{s}}^{LD}(t)) \) and the average value of the ratio \( {\rm 
R}_{L_{s}}^{DL/DD}(t) \)
(\( {\rm R}_{L_{s}}^{LD/DD}(t) \)) in the plateau region to obtain 
both \( {\cal E} \)
and \( Z^{L} \), the matrix element for the local axial current. 
Similarly,
the method is applied to single smeared operators.

A drawback of the BCM  is that, in practice, we have only few different 
cubes at disposal (essentially only two cubes are really useful, \( 
L_{s}=7-9 \), as suggested by earlier 
lattice studies).  Nevertheless, this method is able to give a 
reasonable isolation of the  lightest state and an accurate 
determination of the binding energy (although the method is less 
efficient  for the determination  of the matrix elements of 
the local axial current). 

In order to improve the accuracy of our analysis, we have also used the \textcolor{red}{Multifit Method} 
which consists in performing a global fit of the data for all cube sizes and 
smearing types by minimizing the total \( \chi ^{2} \)~\cite{ape}. 
Consider, for example, the double-smeared operators. In this case, \( \chi ^{2}_{total} \) is 
defined by
\begin{eqnarray}
\chi ^{2}_{total} & = & \sum _{t=t_{i}}^{t_{f}}\, \left\{ \, \sum 
_{L_{s}=7,9}\, \left( 
\frac{C_{L_{s}}^{LD}(t)_{DATA}-C_{L_{s}}^{LD}(t)}{\sigma 
^{LD}_{L_{S}}(t)}\right) \right. \nonumber \\
 & + & \left. \sum _{L_{s}=7,9}\, \left( 
\frac{C_{L_{s}}^{DD}(t)_{DATA}-C_{L_{s}}^{DD}(t)}{\sigma 
^{DD}_{L_{S}}(t)}\right) ^{2}\right\} 
\end{eqnarray}
 where \( \sigma _{L_{s}}^{LD(DD)} \) is the jacknife error of the 
data points.
We also impose the consistency condition that the binding energy \( 
{\cal E} \)
should be the same for all smearing types and cubes sizes. Moreover, 
in order to reduce the effect from excited states, it is convenient to fit the 
data by including the contribution of at least one excited state  to  \( C^{RR'}_{L_{s}}(t) \). 

\section{Strategy of the unquenched analysis.}

Before giving  details on  the simulation parameters, the  calibration of the lattice spacing  and the 
extraction of the binding energy ${\cal E}$, we  discuss the general strategy 
followed in  the analysis  of the unquenched data. This is a 
crucial  issue  due to  the confusion which, we think, exists  in  the literature. 
\par Our main point, which is justified 
below, is  that  \textcolor{red}{the 
correct procedure consists in performing  independent quenched-like  
calculations of \textbf{all}  quantities for  each fixed value of \( k_{sea} \), including 
the spectroscopy, the calibration of the lattice spacing \( a \) and 
the calculation of the relevant matrix elements~\cite{pisa}. Only when all 
the quantities are expressed in physical units, the  results can be extrapolated 
in the sea quark mass.}  Thus, one ends up with a different set of lattice 
parameters for each \( k_{sea} \), such as  the critical kappa, \( k_{cr} \), the 
light quark masses \( k_{u} \) and \( k_{s} \), the lattice spacing and so on. 
Extrapolations in the valence quark mass, \( k_{v} \), should also be performed 
\textbf{at fixed} \( k_{sea} \), without  ever mixing  up the 
extrapolation in the valence and sea quark  masses, which must remain 
distinct steps of the analysis. In all respects, the value of   \( k_{sea} 
\) is an external ``field'' which controls the link dynamics. 
\par The argument is the following. A change in the value of the sea quark 
mass(es) modifies the value of the effective coupling constant, because the latter receives 
contributions from virtual-quark loops. A  change of the coupling constant may induce a rapid variation of the value of 
the lattice spacing which depends exponentially on $\alpha_{s}$. Therefore, strictly speaking, lattice results 
for different  \( k_{sea} \) correspond  to different lattice dynamics  and  are not directly 
comparable. Only when the results  have been converted to physical units, by using the lattice 
spacing extracted for each \( k_{sea} \), comparisons and extrapolations are 
possible. We stress again that a combined (in valence and sea 
quark  masses) chiral extrapolation of lattice quantities, as for example 
the quark masses, may produce incorrect results because in this way we are mixing 
results corresponding to different values of the lattice spacing and all the parameters of the 
extrapolation   do depend on  \( k_{sea} \) through $a$.
\vskip 0.2 cm

\begin{table}
\caption{\label{taulaparam} {\it Simulation parameters of our run to 
extract the binding
energy \protect\( {\cal E}\protect \). The values of the lattice 
spacing, \protect\( a\protect \), and the critical ($k_{cr}$), light 
($k_{u}$) and strange ($k_{s}$) Wilson parameters
 are given separately for  each sea quark mass (corresponding to 
 $k_{sea}$)(see text).}}
\vspace{0.3cm}
{\centering \begin{tabular}{|c|c||c|c|}
\hline 
\multicolumn{4}{|c|}{\textbf{Simulation Parameters}}\\
\hline 
\hline 
\( \beta =5.6 \)&
\( V=24^{3}\times 40 \)&
\( n_{f}=2 \)&
\( N_{conf}=60 \)\\
\hline 
\hline 
\( k_{v} \)&
\multicolumn{3}{|c|}{\( 0.1560,\, 0.1570,\, 0.1575,\, 0.1580 \) }\\
\hline 
\hline 
\multicolumn{2}{|c||}{\( k_{sea}=0.1575 \) }&
\multicolumn{2}{|c|}{\( k_{sea}=0.1580 \)}\\
\hline 
\multicolumn{2}{|c||}{\( a^{-1}=2.51(6) \) GeV}&
\multicolumn{2}{|c|}{ \( a^{-1}=2.54(6) \) GeV}\\
\hline 
\multicolumn{2}{|c||}{\( k_{cr}=0.15927(5) \)}&
\multicolumn{2}{|c|}{\( k_{cr}=0.15887(4) \)}\\
\hline 
\multicolumn{2}{|c||}{\( k_{u}=0.15920(5) \)}&
\multicolumn{2}{|c|}{\( k_{u}=0.15880(4) \)}\\
\hline 
\multicolumn{2}{|c||}{\( k_{s}=0.15747(12) \)}&
\multicolumn{2}{|c|}{\( k_{s}=0.15715(8) \)}\\
\hline 
\end{tabular}\par}\vspace{0.3cm}
\end{table}
 
Having explained  our strategy, we turn to the numerical  results for the binding 
energy. We have performed an unquenched computation of \( {\cal E} \) with 
two degenerate sea quarks at two values of their mass, \( k_{sea}=0.1575 \) and \( 
k_{sea}=0.1580 \).  The heavy and light quark
propagators have been computed  using the set of  unquenched link configurations 
generated by the T\( \chi  \)L Collaboration. Details of the simulation  can be
found in ref.~\cite{lip}.
Light quarks are simulated using the Wilson action whereas heavy 
quarks are described with the discretized HQET. The parameters of our run are 
given in  Table \ref{taulaparam}.  The calibration of the lattice 
spacing has been performed using the $K^{*}-K$ lattice-plane method of 
ref.~\cite{giustileo}.

With the BC method,  we find that the flattest and largest plateaus 
for \( \Delta E^{RR'}_{L_{s}}(t) \) and \( 
{\rm R}_{L_{s}}^{RR'/NN'}(t) \) correspond to the cube size \( L_{s}=7 \) and the 
correlation functions \( LD \) (smearing in the origin) and \( DD \). 
With the Multifit Method, we also obtain that a two-state global fit of the correlations \( 
LD \) and \( DD \) describes very well the data.  The agreement for 
single smeared interpolating operators is, instead, much  worse.  Therefore we base our 
results on  double smeared operators since this is the most efficient 
way of  isolating the lightest state. Our best estimates of the values of the binding energy 
\( {\cal E} \) are
\begin{eqnarray}
a{\cal E}_{B_{d}} & = & \{\, 0.588(11)(5),\, 0.606(15)(2)\, 
\}\nonumber \\
a{\cal E}_{B_{s}} & = & \{\, 0.620(8)(5),\, 0.632(12)(2)\, 
\}\label{eq:valuesofE} 
\end{eqnarray}
for the two values of \( k_{sea}=0.1575 \) and \( 0.1580 \), 
respectively.
The first error is statistical and the second systematic. The latter 
has been obtained from the spread of our results due to different time 
intervals of the fit, cube sizes and smearing types. Since a full 
account of the different methods and evaluation of the uncertainties can be found in ref.~\cite{ape}, we 
do not give further details here.
\section{Sources of systematic error.}
Using eqs.~(\ref{eq:masterformulaallorders}), (\ref{eq:valorsdeD}), (\ref{eq:resultsforX}) 
and (\ref{eq:valuesofE}), we can readily obtain the value of the \( b \) quark mass. As the 
value of \( X_{2} \) is still preliminary and incomplete, we derive our results 
with the two-loop formula corresponding to the NNLO  matching. In order to 
evaluate  the systematic errors on these results, we  carefully studied the 
different sources of uncertainties coming from the use of  eq.~(\ref{eq:masterformulaallorders})
at this order: the value of $\alpha_{s}$, higher-order perturbative 
corrections, input meson mass, method of extracting ${\cal E}$, value of \( 
k_{sea} \) and $1/m_{b}$ corrections.  In the following, unless stated otherwise,
the central values in the tables correspond to \( k_{sea}=0.1580 \), 
\( {\cal E} \) has been obtained with the Multifit Method, the input  mass is 
 the \( B_{s} \) meson mass and a linear chiral extrapolation in \( 
k_{v} \) to \( k_{s} \)  has been performed. In the tables, the first error is statistical 
and the second the systematic one based on the spread of the results due to 
the uncertainty in the lattice spacing, the different forms of writing 
eq.~(\ref{eq:masterformulaallorders}) (see sect. 5.2) and the use of different  time-intervals and fitting 
methods.

\subsection{Dependence of \protect\( 
\overline{m}_{b}(\overline{m}_{b})\protect \) on
\protect\( \alpha _{s}\protect \).}

In order to  obtain \( \overline{m}_{b}(\overline{m}_{b}) \)  we have to 
choose  the value of \( \alpha _{s}\) to be used in the perturbative calculations. 
At all orders, to cancel the renormalon singularities in 
eq.~(\ref{eq:masterformulaallorders}), the same coupling constant  has to be 
used for the lattice and  continuum series. For this reason,   although 
our calculation is  truncated  at ${\cal O}(\alpha^{2}_{s})$,  we  prefer 
to  take the same coupling  constant
for both the lattice and the continuum cases. 
\par One possibility is to consider the 
physical  value of \( \alpha _{s}\) obtained by running the experimental 
coupling   \( \alpha _{s}(M_{Z})=0.118 \) down to \( \alpha _{s}(m_{b}) 
\) with   \( n_{f}=5 \). 
\par A second possibility is to  
account that our simulation has been performed with \( n_{f}=2 \) and 
compute   \( \alpha _{s} \) at the NLO   with a (still to be 
determined)  \( \Lambda_{QCD}^{n_{f}=2} \). In the quenched case, the value of 
\( \Lambda_{QCD}^{n_{f}=0} \sim 250 \) MeV   has  been measured in ref.~\cite{luscher}
Since  the physical value of  \( \Lambda_{QCD}  \) is expected to be 
larger than the quenched one,  in the second case 
we  have  used the NLO value of $\alpha_{s}(m_{b})$ obtained by varying  
 \( \Lambda _{QCD}^{n_{f}=2} \) in the range \( [250,\, 350] \) MeV.  
\par The results for different values of  \( \alpha _{s} \) are presented 
in Table \ref{tabledepalpha}. The dependence on \( \Lambda _{QCD}^{n_{f}=2} \) is very weak: the 
maximum spread of the values is less than \( 20 \) MeV. The difference between the 
central values obtained with the \( n_{f}=5 \) and \( n_{f}=2 \) coupling 
 is of  about \( 50 \) MeV. We have taken this as a very 
conservative estimate of the error due to the choice of the coupling 
constant. \par  
Note that the value of $\alpha_{s}(m_{b})$  for  \( \Lambda _{QCD}^{n_{f}=2}=300\) MeV, 
$\alpha_{s}(m_{b})=0.182$, corresponds with a very good approximation  
to the arithmetic (and geometric) average of the  quenched ($n_{f}=0$) and physical 
($n_{f}=5$) couplings ($\alpha_{s}(m_{b})=0.15$ and $\alpha_{s}(m_{b})=0.22$ 
respectively). For this reason our central value for \( 
\overline{m}_{b}(\overline{m}_{b}) \) is that computed with $\alpha_{s}(m_{b})=0.182$.

\begin{table}

\caption{\label{tabledepalpha} {\it Dependence of \protect\( 
\overline{m}_{b}(\overline{m}_{b})\protect \)
on the choice of \protect\( \alpha _{s}\protect \). For 
$n_{f}=2$ the value of \( \Lambda _{QCD}\) is explicitely given; 
the label $M_{Z}$ indicates that the physical value of 
$\alpha_{S}(M_{Z})$ has been used as input to compute \( \alpha _{s}(\overline{m}_{b}) \) 
 (see text).}}
\vspace{0.3cm}
{\centering \begin{tabular}{|c||c|c||c|}
\hline 
\multicolumn{4}{|c|}{\textbf{Dependence on} \( \alpha _{s} \)}\\
\hline 
\hline 
\( \Lambda _{QCD}\) & \( n_{f} \)&
\( \alpha _{s}(\overline{m}_{b}) \)&
\( \overline{m}_{b}(\overline{m}_{b}) \)\\
\hline 
\( M_{Z}\) &\( 5 \)&
\( 0.221 \)&
\( 4.31(3)(3) \)\\
\hline 
\( 250\) & \( 2 \)&
\( 0.172 \)&
\( 4.25(3)(3) \)\\
\hline 
\( 300\) & \( 2 \)&
\( 0.182 \)&
\( 4.26(3)(3) \)\\
\hline 
\( 350\) & \( 2 \)&
\( 0.191 \)&
\( 4.27(3)(3) \)\\
\hline 
\end{tabular}\par}\vspace{0.3cm}
\end{table}

\subsection{Dependence of \protect\( 
\overline{m}_{b}(\overline{m}_{b})\protect \) on
higher orders.}
 Eq.~(\ref{eq:masterformulaallorders}), which is used to evaluate \(  \overline{m}_{b}(\overline{m}_{b}) \)
to \( {\cal O}(\alpha _{s}^{2}) \)  consists in the product of two  factors. We can, 
then, organize  the formula including (not expanding) or excluding (expanding to \( 
{\cal O}(\alpha _{s}^{2}) \)) the  \( {\cal O}(\alpha _{s}^{3}) \)  terms arising from the product. We take   
the difference between these (formally equivalent) procedures, as an  estimate of  unknown
higher-order terms in perturbation theory (PT). In Table \ref{tabhigher}, the values of \( \overline{m}_{b}(\overline{m}_{b}) \)
obtained from the expanded and not expanded forms of  eq.~(\ref{eq:masterformulaallorders})
are presented. For  \( n_{f}=2 \)  the dependence of our results on higher orders
is  $30$--$40$ MeV,  for  \( n_{f}=5 \)  the  difference is $60$ MeV. From 
these spreads, we  conclude that a fair estimate of the effect of expanding or not expanding 
eq.~(\ref{eq:masterformulaallorders}) is  \( \sim 50 \) MeV.  
From this estimate we
assume from higher-order terms an error of $\pm 25$ MeV. 
\par As  best estimate of \(  \overline{m}_{b}(\overline{m}_{b}) \), for 
each choice of the value of the coupling constant, we take the average of the results obtained 
with the not-expanded and expanded form of the master formula.  In 
this way we have computed  the central values given in tables~\ref{tabledepalpha}, \ref{tableMBmeson}, 
\ref{tablesmeatype}, \ref{tabksea} and \ref{tabcontlim}. 
\begin{table}
\caption{\label{tabhigher}{\it Dependence of \protect\( 
\overline{m}_{b}(\overline{m}_{b})\protect \)
on higher orders in PT: (I) expanded and not expanded form of the master 
equation; (II) excluding or including preliminary (incomplete) 
\protect\( NNLO\protect \)
contributions.   For  $n_{f}=2$ the value of \( \Lambda _{QCD}\) is explicitely given; 
the label $M_{Z}$ indicates that the physical value of 
$\alpha_{S}(M_{Z})$ has been used as input to compute \( \alpha _{s}(\overline{m}_{b}) \).}}
\vspace{0.3cm}
{\centering \begin{tabular}{|c||c|c||c||c|}
\hline 
\multicolumn{5}{|c|}{\textbf{Dependence on higher orders (I)}}\\
\hline 
\hline 
\( \Lambda _{QCD}\) & \( n_{f} \)&
\( \alpha _{s}(\overline{m}_{b}) \)&
\( \overline{m}_{b}(\overline{m}_{b}) \) {\it not expanded}&
\( \overline{m}_{b}(\overline{m}_{b}) \) {\it  expanded} \\
\hline 
\( M_{Z}\) & \( 5 \)&
\( 0.221 \)&
\( 4.28(3)(1) \)&
\( 4.34(3)(1) \)\\
\hline 
\( 300 \) & \( 2 \)&
\( 0.182 \)&
\( 4.24(3)(1) \)&
\( 4.28(3)(1) \)\\
\hline 
\hline 
\multicolumn{5}{|c|}{\textbf{Dependence on higher orders (II)}}\\
\hline 
\hline 
\( \Lambda _{QCD}\) & \( n_{f} \)&
\( \alpha _{s}(\overline{m}_{b}) \)& \( \overline{m}_{b}(\overline{m}_{b}) \)
at \( {\cal O}(\alpha _{s}^{2}) \)&
\( \overline{m}_{b}(\overline{m}_{b}) \)  at \( {\cal O}(\alpha _{s}^{3}) \)\\
\hline 
\( M_{Z}\) & \(5 \)&
\( 0.221 \)&
\( 4.34(3)(1) \)&
\( 4.16 \)--\(4.34 \)\\
\hline 
\( 300\) & \( 2 \)&
\( 0.182 \)&
\( 4.28(3)(1) \)&
\( 4.10 \)--\(4.27 \)\\
\hline \hline
\end{tabular}\par}\vspace{0.3cm}
\end{table}

In order to get an independent estimate of the systematic uncertainty 
due to higher-order terms in the perturbative espansion,  we also tried
to compute  \(  \overline{m}_{b}(\overline{m}_{b}) \)
using   for \( \overline{X}_{2} \)  the preliminary result 
of eq.~(\ref{eq:resultsforX}). In the numerical calculations,
obtained with the expanded form of eq.~(\ref{eq:masterformulaallorders}), 
we allowed \(\overline{X}_2 \) to vary 
in the \( 1\sigma  \) interval \( [64.3,\, 82.7] \), obtaining the range
of masses given in  Table \ref{tabhigher} (only the central values are given). 
The values of \( \overline{m}_{b}(\overline{m}_{b}) \) 
 at order \( {\cal O}(\alpha _{s}^{2}) \)  are also  given for comparison. 
There are  huge numerical cancellations occurring 
in the calculation of $X_2$ from \( \overline{X}_{2} \) 
in  eq.~(\ref{eq:resultsforX}).
For this reason,  the difference between the NNLO and the (approximate)
N$^{3}$LO results 
varies  from about zero to $180$ MeV, depending on the value
of  \( \overline{X}_{2} \). This quantity, even  in the quenched case, is 
still affected by such a large   uncertainty~\cite{direnzo99} that it is impossible
to use it for any realistic estimate. We urgently call for a more precise 
determination of \( \overline{X}_{2} \) in both the unquenched and quenched cases.

\subsection{Dependence of \protect\( 
\overline{m}_{b}(\overline{m}_{b})\protect \) on
the input \protect\( B\protect \) meson mass.}

Consistent values of \( \overline{m}_{b}(\overline{m}_{b}) \) should 
be obtained using as input either the \( B_{d} \) or the 
 \( B_{s} \)  meson masses. The corresponding values of the binding energy, \( {\cal E}_{B_{d}} \) 
and \( {\cal E}_{B_{s}} \), respectively, are  used in the two cases. 
This  checks the lattice value of the $M_{B_{d}}$-$M_{B_{s}}$ 
mass splitting and the  smoothness of  our chiral extrapolation.  
Actually,  for the \( B_{s} \) meson, the  physical value of  the strange quark mass, corresponding  
to \( k_{s} \), is within the range of  valence quark masses  (see Table \ref{taulaparam}) and only a mild 
interpolation, rather than an extrapolation, is needed. For
the \( B_{d} \) meson, instead, we have extrapolated  almost to the chiral limit. 
In order to compute the pole mass we have taken 
\( M_{B_{d}}=5.279 \) and \( M_{B_{s}}=5.375 \) GeV~\cite{pdg}. In 
Table \ref{tableMBmeson}, we compare our results for the two cases. 
The results are nicely compatible and a  small difference 
($\simeq 30$ MeV) is observed  in the two cases. 
\begin{table}
\caption{\label{tableMBmeson}{\it Dependence of \protect\( 
\overline{m}_{b}(\overline{m}_{b})\protect \)
on the input \protect\( B\protect \) mesons masses (see text).}}
\vspace{0.3cm}
{\centering \begin{tabular}{|c||c|c||c||c|}
\hline 
\multicolumn{5}{|c|}{\textbf{Dependence on} \( M_{B} \) }\\
\hline 
\hline 
\( \Lambda _{QCD}\) & \( n_{f} \)&
\( \alpha _{s}(\overline{m}_{b}) \)&
\( \overline{m}_{b}(\overline{m}_{b}) \) from \( B_{d} \)&
\( \overline{m}_{b}(\overline{m}_{b}) \) from \( B_{s} \)\\
\hline 
\( M_{Z}\) & \(5 \)&
\( 0.221 \)&
\( 4.28(4)(3) \)&
\( 4.31(3)(3) \)\\
\hline 
\( 300 \) & \(2 \)&
\( 0.182 \)&
\( 4.23(4)(3) \)&
\( 4.26(3)(3) \)\\
\hline \hline
\end{tabular}\par}\vspace{0.3cm}
\end{table}
 
\subsection{Dependence of \protect\( 
\overline{m}_{b}(\overline{m}_{b})\protect \) on
the method for extracting \protect\( {\cal E}\protect \).}

In sect. 3, we discussed the two methods  used to determine 
the binding energy \( {\cal E} \): the Best Cube method and the Multifit method. 
If the lightest state has been well isolated, both methods should give 
compatible results for the \( b \) quark mass. In Table \ref{tablesmeatype}, we present 
the value of \( \overline{m}_{b}(\overline{m}_{b}) \) for the BC method 
(obtained with \( LD/DD \), smearing in the origin, and \( L_{s}=7 \), which gives 
the flattest and largest plateau) and for the Multifit method.  Also 
in this case the results differ by $\simeq 30$ MeV.
\begin{table}

\caption{\label{tablesmeatype}{\it Dependence of \protect\( 
\overline{m}_{b}(\overline{m}_{b})\protect \)
on the method to extract \protect\( {\cal E}\protect \) (see text).}}
\vspace{0.3cm}
{\centering \begin{tabular}{|c|c|c||c||c|}
\hline 
\multicolumn{5}{|c|}{\textbf{Dependence on the method of extracting 
\( {\cal E} \)}}\\
\hline 
\hline 
\( \Lambda _{QCD}\) & \(n_{f} \)&
\( \alpha _{s}(\overline{m}_{b}) \)&
\( LD/DD \) with \( L_{s}=7 \)&
Multifit\\
\hline 
\( M_{Z}\) & \(5 \)&
\( 0.221 \)&
\( 4.34(2)(4) \)&
\( 4.31(3)(3) \)\\
\hline 
\( 300\) & \(2 \)&
\( 0.182 \)&
\( 4.29(2)(3) \)&
\( 4.26(3)(3) \)\\
\hline 
\end{tabular}\par}\vspace{0.3cm}
\end{table}

\subsection{Dependence of \protect\( 
\overline{m}_{b}(\overline{m}_{b})\protect \) on
\protect\( k_{sea}\protect \).}

As discussed before, in order to compare the values  obtained
for different \( k_{sea} \) and attempt an extrapolation, we have 
first to  convert the lattice quantities to physical units. In Table \ref{tabksea}, the 
values of \( \overline{m}_{b}(\overline{m}_{b}) \) for either values 
of \( k_{sea} \)  are given.
The dependence of our results on \( k_{sea} \), in  the sea-quark mass region of our simulation, 
is  small. Indeed, taking into account that they correspond to independent 
simulations,  we are not able to observe  any    dependence
on \( k_{sea} \) within errors. 
Therefore, the only possible  strategy is not to attempt an extrapolation in  \( 
k_{sea} \) and take  the value at the lightest \( k_{sea} \), i.e. \( 
k_{sea}=0.1580 \), as the best  estimate of the physical value of the \( b \) quark mass.  
The difference between the results at the two values of \( 
k_{sea} \) is  accounted as a systematic effect in the final error.

\begin{table}
\caption{\label{tabksea}{\it Dependence of \protect\( 
\overline{m}_{b}(\overline{m}_{b})\protect \)
on  \protect\( k_{sea}\protect \)(see 
text).}}
\vspace{0.3cm}
{\centering \begin{tabular}{|c||c|c||c||c|}
\hline 
\multicolumn{5}{|c|}{\textbf{Dependence on} \( k_{sea} \) }\\
\hline 
\hline 
\( \Lambda _{QCD}\)& \( n_{f} \)&
\( \alpha _{s}(\overline{m}_{b}) \)&
\( \overline{m}_{b}(\overline{m}_{b}) \) with  \( k_{sea}=0.1575 \)&
\( \overline{m}_{b}(\overline{m}_{b}) \) with \( k_{sea}=0.1580 \)\\
\hline 
\( M_{Z}/\) & \( 5 \)&
\( 0.221 \)&
\( 4.34(2)(4) \)&
\( 4.31(3)(3) \)\\
\hline 
\( 300\) & \(2 \)&
\( 0.182 \)&
\( 4.29(2)(4) \)&
\( 4.26(3)(3) \)\\
\hline 
\hline 
\end{tabular}\par}\vspace{0.3cm}
\end{table}

\subsection{\protect\( 1/m_{b}\protect \) correcctions to \protect\( 
\overline{m}_{b}(\overline{m}_{b})\protect \).}

Our whole analysis is performed to the lowest 
order of  the expansion  in \( 1/m_{b} \). This means that \( {\cal O}(1/m_{b}) \) 
contributions to the relation between the QCD and the lattice HQET quantities in
eqs.~(\ref{eq:matching}) and  (\ref{eq:hqetmassformula}) have been  neglected.
We now make an estimate of the error introduced by these higher-order 
corrections.

The HQET pseudoscalar mass formula including \( 1/m_{b} \) 
corrections is given by 
\begin{equation}
\label{eq:haetmass1/m}
M_{B}\, =\, m_{b}\, +\, {\cal E}\, -\, \frac{\lambda _{1}}{2\, 
m_{b}}\, -\, \frac{3\, \lambda _{2}}{2\, m_{b}}\, +\, {\cal 
O}(1/m^{2}_{b}) \, , 
\end{equation}
where the parameters \( \lambda _{1} \) and \( \lambda _{2} \) are  matrix
elements between \( B \)  states of the kinetic and  chromomagnetic operators
\begin{eqnarray}
\lambda _{1} & \equiv  & \frac{\langle B|\, \bar{b}\, \vec{D}^{2}\, 
b\, |B\rangle }{2\, M_{B}}\, , \nonumber \\
\lambda _{2} & \equiv  & \frac{\langle B|\, \bar{b}\, (\vec{S}\cdot 
g\vec{B})\, b\, |B\rangle }{M_{B}}\, , \label{eq:deflam12} 
\end{eqnarray}
with \( \vec{S} \) the spin operator of the \textbf{\( B \)}-meson 
and \( \vec{B} \)
the chromomagnetic field (\( B_{i}=\frac{1}{2}\, \epsilon 
_{ijk}\, G_{jk} \)).
\par It is straightforward  to estimate \( \lambda _{2} \) because 
which is related  to the vector-pseudoscalar  mass splitting,
\begin{equation}
\label{eq:valuelam2}
\lambda _{2}=\frac{1}{4}\, \left( M^{2}_{B^{*}}-M^{2}_{B}\right) \, 
\approx \, 0.12\, GeV^{2}
\end{equation}

The extraction of \( \lambda _{1} \), is, instead, more  difficult 
as demonstrated by the spread of  values obtained with different 
approaches: the lepton-energy spectrum in inclusive 
semileptonic \( B \) decays using Zero Recoil sum rules, QCD Sum rules, experiment 
data analysis and the HQET Virial Theorem (see~\cite{guidoreview} and 
references therein). It has also been estimated on the lattice using the 
discretized HQET~\cite{oursmb,ourlambda1}. Although biased by 
the lattice results, we prefer a small value for this parameter, in 
the absence of an accurate determination  we let it to vary in the 
interval $-0.5$--$0.0$ GeV$^{2}$.
With this range, we find that the contribution of  the \( 1/m_{b} \) 
corrections to the pole mass is at most \( \simeq 30 \) MeV. 
 Due to the theoretical uncertainties on \( \lambda _{1} \), we do not attempt 
to correct   the \( 1/m_{b} \) terms but 
include their effect as a sytematic error on the final result.

\subsection{Continuum limit of \protect\( 
\overline{m}_{b}(\overline{m}_{b})\protect \).}

To date, in  quenched lattice
simulations the binding energy \( a{\cal E} \) has been computed at three values of
 \( \beta =6.0 \), \( 6.2 \) and \( 6.4 \)~\cite{ape}~\footnote{ 
Finite volume effects  may be present in the  results at \(\beta=6.4\) 
since the lattice  volume  was rather  small.}. In Table \ref{tabcontlim} 
we give the  values of the \( b \)-quark mass from these quenched  simulations 
for  different lattice spacings. The quenched results are computed   using
the values of the binding energy from the APE Collaboration~\cite{ape} and
 NNLO quenched master formula  with  
the  coupling constant \( \alpha _{s}(\overline{m}_{b})=0.15 \). The quenched
values are very close to  our new result with  \( n_{f}=2\). 
Although one may argue that there is a (rather mild) tendency towards lower values 
as $a$ decreases,   with the present uncertainties we cannot attempt any 
extrapolation in $a$ or  realistic estimate of the discretization 
errors. 
\begin{table}
\caption{\label{tabcontlim}{\it Results for \protect\( 
\overline{m}_{b}(\overline{m}_{b})\protect \)
at different values of the lattice spacing \protect\( a\protect \). 
The first
error is due to the uncertainties in the lattice determination of the 
binding
energy, and the second is an estimate of the higher order 
perturbative corrections.}}
\vspace{0.3cm}
{\centering \begin{tabular}{|c||c||c||c|}
\hline 
\multicolumn{4}{|c|}{\textbf{Dependence on \( a \) in the quenched 
case}}\\
\hline 
\hline 
\( \beta  \)&
\( a^{-1}(GeV) \)&
\( a{\cal E}_{B_{d}} \)&
\( \overline{m}_{b}(\overline{m}_{b}) \)\\
\hline 
\( 6.0 \)&
\( 2.0(2) \)&
\( 0.61(1) \)&
\( 4.34(5)(10) \)\\
\hline 
\( 6.2 \)&
\( 2.9(3) \)&
\( 0.52(1) \)&
\( 4.29(7)(10) \)\\
\hline 
\( 6.4 \)&
\( 3.8(3) \)&
\( 0.460(7) \)&
\( 4.25(7)(10) \)\\
\hline 
\end{tabular}\par}\vspace{0.3cm}
\end{table}

\section{Final result for \protect\( 
\overline{m}_{b}(\overline{m}_{b})\protect \)
and comparison with other determinations.}

We consider as  best estimate of \( 
\overline{m}_{b}(\overline{m}_{b}) \)
the value obtained  with the pole mass extracted by using the mass of the \( B_{s} \) 
meson and the binding energy \( {\cal  E}_{B_{s}} \)   
measured on the lattice through the Multifit method at \( 
k_{sea}=0.1580 \),  by averaging the results of the expanded and not 
expanded form of eq.~(\ref{eq:masterformulaallorders}) and  by taking the  
NLO  coupling constant \( \alpha _{s} \) computed at NLO with \( n_{f}=2 \) and
 \( \Lambda _{QCD}^{n_{f}=2}=300 \)  MeV.   Using the estimate of the 
 different errors discussed in the previous section we then obtain:
\textcolor{red}{ 
\begin{equation}
\label{eq:resultmb}
\overline{m}_{b}(\overline{m}_{b})\, =\, (\, 4.26\, \pm \, 0.03\, \pm 
0.05\, \pm \, 0.07\, )\, {\rm GeV} \, .
\end{equation}}
The first error is statistical. The second is the systematic 
error obtained from the spread of values due to the use of different time intervals, 
fitting methods, smearing types and cube sizes for the interpolating operators, the 
dependence of the results on the \( k_{sea} \), 
the calibration of the lattice spacing and the estimate of the \( 1/m_{b} \) corrections. 
Finally, the third is an estimate of the error due to the uncertainties in the values 
of \( \alpha _{s} \)  and to the effects of higher-order terms  
in eq.~(\ref{eq:masterformulaallorders}). We find  that the latter is 
the most important source of error in the final result. For this 
reason a  big effort must be done to compute the  unknown N$^{3}$LO contributions  to 
the residual mass on the lattice and the NNLO matching coefficient 
between  the lattice  and  continuum $\alpha_{s}$ in the unquenched 
case. On the numerical side, a non-perturbative calculation  of  \( \Lambda _{QCD}^{n_{f}=2} \) 
is also important.
\par Our new result (\ref{eq:resultmb}) modifies and improve the previous 
one obtained from quenched lattice simulations with  NLO  matching 
only~\cite{oursmb}
\begin{equation}
\label{eq:oldmb}
\overline{m}_{b}(\overline{m}_{b})\, =\, (\, 4.15\, \pm \, 0.05\, \pm 
0.20\, )\, {\rm GeV}
\end{equation}
where the first error is due to the lattice systematics and the  second is an
estimate of higher orders.
\par
\begin{figure}
\begin{center}
\leavevmode
{\resizebox*{0.85\textwidth}{0.6\textheight}{\epsfbox{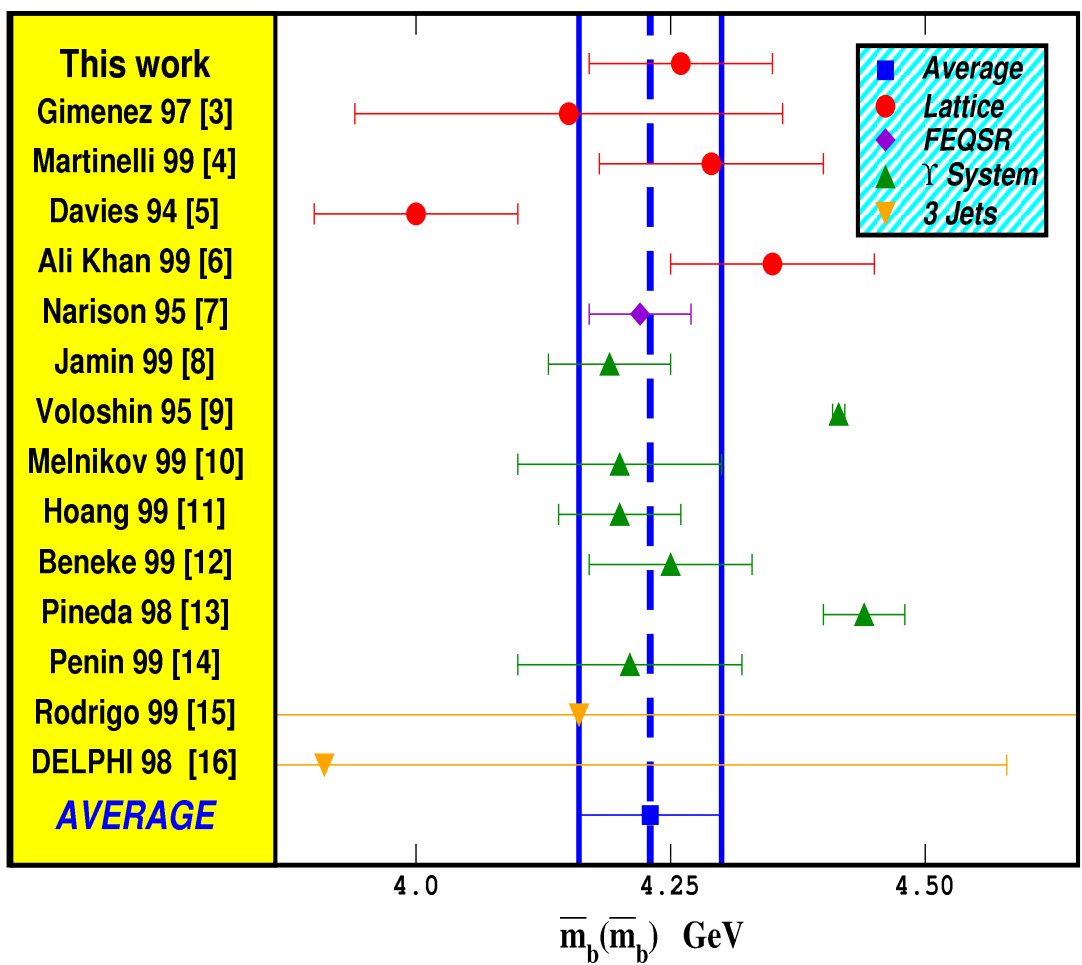}}}
\end{center}
\caption{\it{ Determinations of \( 
\overline{m}_{b}( \overline{m}_{b}) \) from different 
methods compared with our result. Only
references after 1994 have been included.}}
\label{fig:mbcomp}
\end{figure}

 It is interesting to compare  eq.~(\ref {eq:resultmb})
with   recent values  obtained with completely different approaches  as mass effects in 3-jets
 \( b\bar{b}g  \) events, \( b\bar{b} \) production cross-section and \( 
 \Upsilon  \) spectroscopy. Our final result is in good agreement with most of NNLO estimates, as 
shown  in fig.\ref{fig:mbcomp}. In the figure we also give our world  
average and error. This average has been obtained by using only the 
most recent NNLO determinations  from \(   \Upsilon  \) spectroscopy 
and lattice QCD,  i.e. we did not use the results of 
refs.~\cite{oursmb,nrqcd,jamin,voloshin}, either because they have 
been superseeded by more accurate calculations or because they are only 
computed at the NLO accuracy.  
The average is
\textcolor{red}{\begin{equation} 
\overline{m}_{b}(\overline{m}_{b})\, =\, 4.23 \pm 0.07
\, {\rm GeV} \end{equation} }
which corresponds to a relative error of less than $2 \% $ comparable to the 
precision on the top quark mass. The masses of the quarks of the heaviest 
and last discovered generation  are, and will probably remain,  the 
most accurately determined quark masses. 
\textcolor{blue}{\section*{Acknowlegments}} We are extremely grateful
to all the members of the T\( \chi  \)L Collaboration for providing
us with the gauge configurations necessary to this study.
We thank  our  collaborators
D.~Becirevic and V.~Lubicz  for  illuminating discussions on 
the subject of this paper. V. G. has been supported by CICYT under 
Grant AEN-96-1718, by DGESIC under Grant PB97-1261 and by the Generalitat
Valenciana under Grant GV98-01-80. L. G. has been supported in part under 
DOE Grant DE-FG02-91ER40676. G. M. and F. R. acknowledge the M.U.R.S.T. and the INFN
for partial support.

\end{document}